\documentclass[twoside,twocolumn,9pt]{article}
\usepackage{extsizes}
\usepackage[super,sort&compress,comma]{natbib} 
\usepackage[version=3]{mhchem}
\usepackage[left=1.5cm, right=1.5cm, top=1.785cm, bottom=2.0cm]{geometry}
\usepackage{balance}
\usepackage{mathptmx}
\usepackage{sectsty}
\usepackage{graphicx} 
\usepackage{lastpage}
\usepackage[format=plain,justification=justified,singlelinecheck=false,font={stretch=1.125,small,sf},labelfont=bf,labelsep=space]{caption}
\usepackage{float}
\usepackage{fancyhdr}
\usepackage{fnpos}
\usepackage[english]{babel}
\addto{\captionsenglish}{%
  
}
\usepackage{array}
\usepackage{droidsans}
\usepackage{charter}
\usepackage[T1]{fontenc}
\usepackage[usenames,dvipsnames]{xcolor}
\usepackage{setspace}
\usepackage[compact]{titlesec}
\usepackage{hyperref}

\usepackage{epstopdf}

\usepackage{amssymb} 
\definecolor{cream}{RGB}{222,217,201}

\begin{document}

\pagestyle{fancy}
\thispagestyle{plain}
\fancypagestyle{plain}{
\renewcommand{\headrulewidth}{0pt}
}

\makeFNbottom
\makeatletter
\renewcommand\LARGE{\@setfontsize\LARGE{15pt}{17}}
\renewcommand\Large{\@setfontsize\Large{12pt}{14}}
\renewcommand\large{\@setfontsize\large{10pt}{12}}
\renewcommand\footnotesize{\@setfontsize\footnotesize{7pt}{10}}
\makeatother

\renewcommand{\thefootnote}{\fnsymbol{footnote}}
\renewcommand\footnoterule{\vspace*{1pt}%
\color{cream}\hrule width 3.5in height 0.4pt \color{black}\vspace*{5pt}} 
\setcounter{secnumdepth}{5}

\makeatletter 
\renewcommand\@biblabel[1]{#1}            
\renewcommand\@makefntext[1]%
{\noindent\makebox[0pt][r]{\@thefnmark\,}#1}
\makeatother 
\renewcommand{\figurename}{\small{Fig.}~}
\sectionfont{\sffamily\Large}
\subsectionfont{\normalsize}
\subsubsectionfont{\bf}
\setstretch{1.125} 
\setlength{\skip\footins}{0.8cm}
\setlength{\footnotesep}{0.25cm}
\setlength{\jot}{10pt}
\titlespacing*{\section}{0pt}{4pt}{4pt}
\titlespacing*{\subsection}{0pt}{15pt}{1pt}

\fancyfoot{}
\fancyfoot[LO,RE]{\vspace{-7.1pt}\includegraphics[height=9pt]{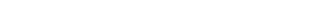}}
\fancyfoot[CO]{\vspace{-7.1pt}\hspace{13.2cm}\includegraphics{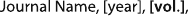}}
\fancyfoot[CE]{\vspace{-7.2pt}\hspace{-14.2cm}\includegraphics{head_foot/RF}}
\fancyfoot[RO]{\footnotesize{\sffamily{1--\pageref{LastPage} ~\textbar  \hspace{2pt}\thepage}}}
\fancyfoot[LE]{\footnotesize{\sffamily{\thepage~\textbar\hspace{3.45cm} 1--\pageref{LastPage}}}}
\fancyhead{}
\renewcommand{\headrulewidth}{0pt} 
\renewcommand{\footrulewidth}{0pt}
\setlength{\arrayrulewidth}{1pt}
\setlength{\columnsep}{6.5mm}
\setlength\bibsep{1pt}

\makeatletter 
\newlength{\figrulesep} 
\setlength{\figrulesep}{0.5\textfloatsep} 

\newcommand{\topfigrule}{\vspace*{-1pt}%
\noindent{\color{cream}\rule[-\figrulesep]{\columnwidth}{1.5pt}} }

\newcommand{\botfigrule}{\vspace*{-2pt}%
\noindent{\color{cream}\rule[\figrulesep]{\columnwidth}{1.5pt}} }

\newcommand{\dblfigrule}{\vspace*{-1pt}%
\noindent{\color{cream}\rule[-\figrulesep]{\textwidth}{1.5pt}} }

\makeatother

\twocolumn[
  \begin{@twocolumnfalse}
{\includegraphics[height=30pt]{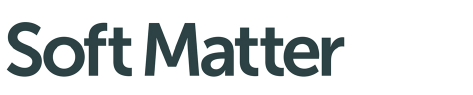}\hfill\raisebox{0pt}[0pt][0pt]{\includegraphics[height=55pt]{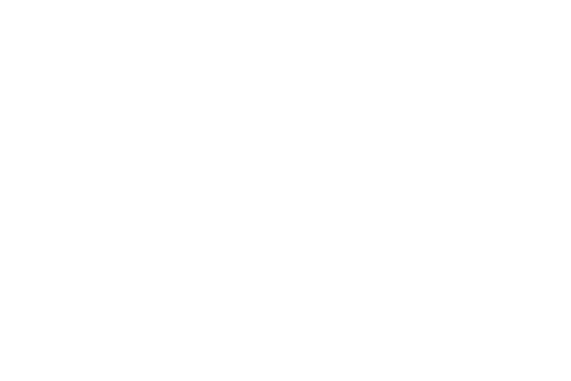}}\\[1ex]
\includegraphics[width=18.5cm]{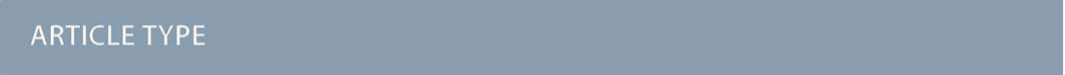}}\par
\vspace{1em}
\sffamily
\begin{tabular}{m{4.5cm} p{13.5cm} }

\includegraphics{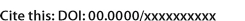} & \noindent\LARGE{\textbf{Peeling threshold for removal of an adhered elastic sheet  by a shear flow}} \\
\vspace{0.3cm} & \vspace{0.3cm} \\

& \noindent\large{Hugo Perrin\textit{$^{a,b}$} and Lorenzo Botto\textit{$^{a}$}} \\

\includegraphics{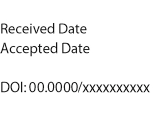} & \noindent\normalsize{Fluid shear can induce detachment of a  thin elastic sheet adhered to  a flat substrate. This peeling process is important in a variety of environmental and technological systems. The condition for peeling depends on: the shear rate $\dot{\gamma}$, the fluid viscosity $\eta$, the length of the detached portion of the sheet $L$, the bending rigidity $B$ and the adhesion energy $\Gamma$. What are the laws governing the detachment? We address this question experimentally in the regime of intermediate adhesion, using macroscopic sheets bonded to a substrate and immersed in a shear cell containing a viscous fluid. The experiments indicate a critical shear rate for peeling of the order of $\dot{\gamma} \sim B/(\eta L^3)$. This threshold is, unexpectedly, independent of adhesion. We rationalise this result by applying Griffith's fracture theory to  optical measurement data of the shape of the sheet, under conditions of freely moving peeling front or clamped boundary. The results indicate that the large curvature of the sheet for $\dot{\gamma} \sim B/(\eta L^3)$ yields a nearly diverging strain energy release rate at this threshold. This approximate divergence in turn yields a  peeling threshold that depends at most weakly on $\Gamma$, confirming a theory that was proposed recently (Salussolia et al., J. Mech. Phys. Solids, 2020, 134). Among other applications, our work provides a quantitative formula that can aid the production at scale of 2D materials such as graphene.   } \\

\end{tabular}

 \end{@twocolumnfalse} \vspace{0.6cm}

  ]

\renewcommand*\rmdefault{bch}\normalfont\upshape
\rmfamily
\section*{}
\vspace{-1cm}


\footnotetext{\textit{$^{a}$~Process and Energy Department, 3ME Faculty of Mechanical, Maritime and Materials Engineering,
TU Delft, 2628 CD Delft, The Netherlands}}
\footnotetext{\textit{$^{b}$Paris-Saclay University, FAST, 91405, Orsay, France}}





\section{Introduction}

The application of fluid shear can produce the detachment of fragments of adhesive materials or particles from a flat substrate immersed in a liquid. This basic process occurs in  a variety of systems. For example, in biology the adhesion and detachment of a cell from the vessel walls during blood flow plays a crucial role in healing.   \cite{DONG200035} In water filtration, biofilm removal can be obtained by flow reversal or increased flow rate. \cite{QU201230}  In microelectronic, the flow-induced removal of contaminant particles is exploited in the cleaning of silicon wafers. \cite{JoR14Walker} Flow-induced detachment by water jets is increasingly used in automatic fruit peeling machines \cite{shi2025optimization} and for skin abrasion for cosmetic or medical uses.   \cite{chen2024comparison} A key  question in many of these applications is the determination of the critical shear rate required to detach the adhered object. This quantity is important not only to engineer ways of removing foreign material, but also for the quantification of the magnitude of the attachment forces from the knowledge that detachment has taken place. 

An important class of adhered materials takes the form of small sheets, i.e., thin plate-like objects. Flow-induced removal of sheets plays a role in the exfoliation of 2D layered materials, \cite{agrawal2022viscous, kamal2021effect, qi2025unraveling, chen2022dynamic, Paton2014ty} flaking of paint due to moving water or wind, \cite{de2005simulation} weathering of stones, flaking of embrittled plastics, \cite{weinstein2016macroplastic} and removal of flakes produced by corrosion.  \cite{effendy2021blistering}  Here the interplay between the flow induced forces, the bending deformation of the sheet, and adhesion controls the probability of detachment. A recent application that calls for an improved understanding of the fluid-solid coupling is the modelling of the liquid-exfoliation method to produce 2D materials such as graphene, hexagonal boron nitride, etc. In this process, a layered material - graphite powder in the case of graphene - is suspended in a liquid solvent and subjected to energetic turbulence. The shear flow experienced by each particle of layered material forces particle delamination, yielding single- or few layer nanomaterial sheets.  \cite{Paton2014ty} For process optimisation, the relation between the critical shear rate,  the lateral size of the sheet, the bending rigidity and the adhesion energy is required. Despite several studies, \cite{kim2021prediction, qi2025unraveling, chen2022dynamic}  the flow-induced delamination process determining 2D material exfoliation is not completely understood. \cite{botto2019toward} 

A recently proposed hydrodynamic peeling model  suggests that a parameter range exists where the critical shear rate $\dot{\gamma}$ depends mostly on the bending rigidity of the sheet $B$ and the length $L$ of the detached portion of the sheet, but weakly on adhesion. \cite{SALUSSOLIA2020103764} The model for the critical shear rate $\dot{\gamma}_c$ has an approximate power-law dependence on $L$ according to the formula 
\begin{equation}\label{eqn:criticalrate}
\dot{\gamma}_c = \frac{B}{\eta L^3} f(\Gamma)
\end{equation}
where $\eta$ is the dynamic viscosity of the fluid, and $f$ is an  $O(1)$ quantity which is a weak function of the adhesion energy $\Gamma$ (where $\Gamma$,  the fracture energy of the solid-solid interface, is  appropriately normalised with the remaining dimensional parameters). The theoretical prediction  (\ref{eqn:criticalrate}) has not yet been verified experimentally. The current paper aims to fill this gap using centimeter sized elastic sheets of controlled bending rigidity, length and adhesion properties. More generally, our investigation provides experimental insights into the critical conditions for peeling and its relation to the shape of the peeled sheet . 

Ours is a contribution to the collective research effort by the mechanics and soft matter communities to understand problems of peeling of sheet-like materials bounded by liquids.  In this area of research, many investigations have focused on the blistering phenomenon, where fluid pressure pushes on a thin sheet that completely encloses the liquid, driving an adhesive failure. \cite{juel2018instabilities, dhong2015coupled,pandey2023hoop,cao2026fracture,ball2018static}The current investigation is distinct from  work on blistering in that the peeling process  is here driven by a shear flow that pushes a flap that is open (the flap forms a wedge whose opening angle increases in time). The physical situation is similar to that of peeling induced by a mechanical manipulator (e.g. by a tweezers or a moving support), \cite{dhong2017peeling, memet2021static} except that in our case the load - which is of the follower type  \cite{barbieri2022peeling,bigoni2023flutter, pramanik2024computational,migliaccio2025viscous} - is produced by pressure and viscous forces that depend on the instantaneous configuration of the sheet (and are not assigned a priori).  The coupling of solid deformation to flow makes theoretical predictions particularly arduous. For example, no simple expressions for the local drag are available for sheets of significant extent in the depth direction, \cite{salussolia2025slender} unlike for fibers, for which the slender body theory can be used locally. \cite{man2025slip} Thus this work addresses a problem that overlaps with fluid-structure interaction subjects that have been treated in the literature, but that are not identical to the problem at hand.

\section{Experimental set-up}
The experiments are carried out in a  shear cell composed of two co-rotating cylinders driving a belt. The parallel translation of the belt imposes a simple shear (Couette) flow with shear rate in the range $\dot \gamma = 0.1- 10\rm{s^{-1}}$. The liquid used is glycerol (viscosity $\eta \simeq 1.3\rm{Pa.s}$, density $\rho \simeq 1.26\times10^{3}\;\rm{kg.m^{-3}}$). 
The inextensible flexible adhesive sheets are made of a silicone elastomer. The silicone elastomer is obtained by a polymerization reaction from monomers in the liquid phase by adding a liquid catalyst. To minimise buoyancy, graphite powder is added to the liquid mixture to approximately match the density of the elastomer with that of the glycerol (density matching was obtained using a $36\%$  fraction in mass of Synthetic Aldrich graphite powder of size < $20\rm{\mu m}$). After mixing, the liquid mixture of monomers, catalyst and graphite powder is blade-coated into films of thickness $t$ ranging from $100 \rm{\mu m}$ to $500 \rm{\mu m}$. The blade coating process must be performed slowly to avoid shear-induced migration of the graphite powder particles (shear-induced migration can lead to a concentration gradient of graphite particles in the film thickness and cause anisotropic elasticity).  After polymerization (20min at room temperature), elastic films of uniform thickness are obtained. 

The Young's modulus of the solidified filled elastomer is $E \simeq 3.7\times 10^{2}\rm\;{kPa}$ (measured with an Anton Paar MCR302 Rheometer). The bending modulus of the sheets $B= t^3 E /12(1-\nu^2)$ varies from $10^{-6}J$ to $10^{-8}\rm{J}$, where $\nu\simeq 0.5$ is the Poisson's ratio. The films were cut into long rectangles of width $w=1\;\rm{cm}$ and total length $10\;\rm{cm}$. The elastomer sheet was then placed on a thin layer of the same elastomer, which was in turn strongly adhered to a glass slide. Care was taken not to trap any bubbles between the elastomer surfaces. The adhesion energy parameter characterising the sheet-sheet interfacial bonding in glycerol is $\Gamma = 2 \gamma_{sheet/gly}$ where $\gamma_{sheet/gly}$ is the surface energy of the sheet with glycerol. To estimate $\Gamma$ we measured the contact angle of a glycerol drop on the  sheet, obtaining $\theta_Y \simeq 85^\circ$. According to the Young-Dupre equation $\gamma_{sheet/gly} = \gamma_{sheet/air} -  \gamma_{gly/air} \cos \theta_Y$ \cite{deGe02}. Assuming  $\gamma_{sheet/air}\simeq \gamma_{silicone/air}\simeq  2  \times 10^{-2}\;\rm{J.m^{-2}}$ and $\gamma_{gly/air}\simeq 6  \times 10^{-2}\;\rm{J.m^{-2}}$ we obtained $\Gamma\simeq 3  \times 10^{-2}\;\rm{J.m^{-2}}$.

The flat substrate with the attached deformable sheet is immersed in the shear cell. One end of the tape is kept detached with the help of a tweezer. Then a constant shear rate is imposed. Optical imaging is used to measure the shape of the tape and the instantaneous length of the detached portion of the tape. The pixel sizes of the images are $1920\times1080$, the spatial resolution is order of $25\;\rm{\mu m/pix}$.
\begin{figure}
\centerline{\includegraphics{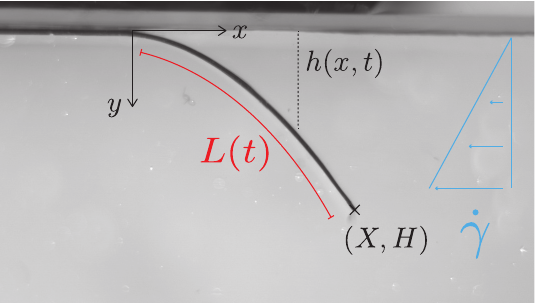}}
\caption{Side view image of the experiment. An adhesive elastomeric sheet (curved line) is adhered to a rigid substrate (top), a plexiglass plate covered with an adhesive elastomeric sheet. The length of the non-adhered part is $L$. The sheet is subjected to a simple shear flow of shear rate $\dot \gamma$. In a coordinate system $(x,y)$ centered at the peeling front, the coordinates of the tip of the sheet are $x=X$ and $y=H$. The equation describing the shape of the sheet is $y=h(x,t)$. We seek to understand the shape of the sheet as a function of $L$, $\dot{\gamma}$ and the bending rigidity $B$ of the sheet, as well as the critical combination of these parameters that produces motion of the peeling front.  }
\label{fig1}
\end{figure}

\section{Results}

\begin{figure}
\centerline{\includegraphics[width = 9cm]{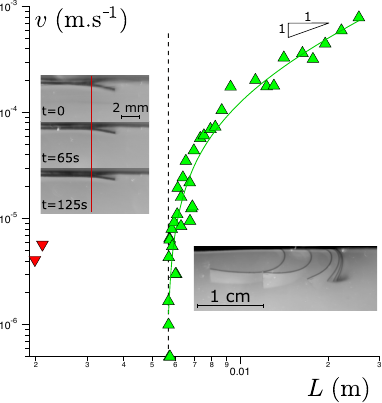}}
\caption{Peeling front velocity versus $L$ for $\dot \gamma \simeq 4.1\rm{s^{-1}}$ and $B\simeq4.6 \times 10^{-7}J$. Plots such as this are used to obtain the critical value of $L$. The green and red markers indicate a positive velocity (peeling) and a negative velocity (reattaching), respectively. The green line is  $v = v_0 (L/L_c - A)^\beta$, with $L_c = (B/\eta \dot \gamma )^{1/3}\simeq 4.4\;\rm{mm}$, $v_0\simeq 1.2\times10^{-4}\;\rm{m.s^{-1}}$, $A\simeq1.2 $ and $\beta \simeq 1.2$. The three inset images on the left illustrate the reattachment dynamics, the vertical red line marking the position of the front at $t=0$.  The inset image on the right shows the shape of a sheet for $L$ larger than the critical peeling value, for times $t=0s, 48s, 90s, 150s, 190s,210s$ .}
\label{figA}
\end{figure}

We performed dynamic peeling experiments by fixing the shear rate $\dot \gamma $ and, for a given bending modulus $B$, varying the initial length of the non-adhered portion of the sheet, $L$. Above a critical value of $L$ the sheet peels off, the non-adhered portion of the sheet increases in length and bends towards the flow direction, see the bottom right caption of figure \ref{figA}. For values of $L$ smaller than the critical value the sheet reattaches, see the top left captions of figure \ref{figA}. Plotting the velocity of the peeling front vs. $L$ enabled us to measure with precision the critical value of $L$ separating peeling from reattachment (for a given value of $\dot{\gamma}$ and $B$). For example, in Fig. \ref{fig1} we see that the critical value is $L\simeq 5.8\;\rm{mm}$.  

\begin{figure}[h!]
\centerline{\includegraphics[width = 9cm]{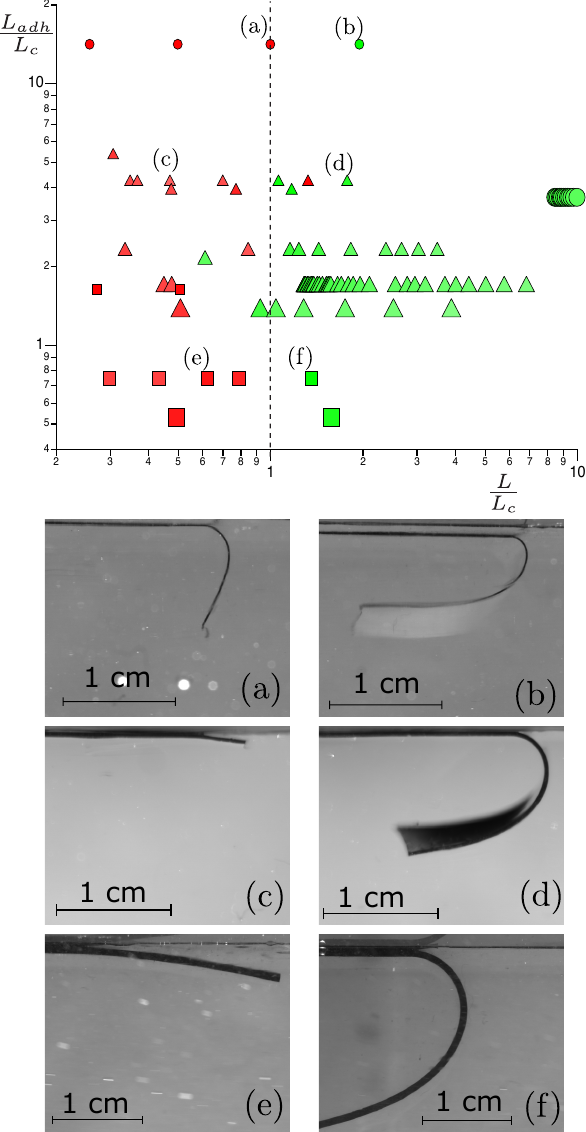}}
\caption{Phase diagram mapping reattachment or peeling regions in the  $L/L_c$ and $L_{adh}/L_c$ space, where $L_c = (B/\eta \dot \gamma )^{1/3}$and  $L_{adh}= \Gamma / (\eta \dot \gamma)$ . Red markers indicate reattachment, green markers indicate peeling. The sizes of the markers corresponds to the magnitude of the shear rates, from $0.7$ to $5.7\;\rm{s^{-1}}$ (larger symbols for larger shear rates). The marker types corresponds to different values of the bending modulus $\circ$  : $B \simeq 5.8 \times 10^{-8}\;\rm{J}$,$\triangle$  : $B \simeq 4.6 \times 10^{-7}\;\rm{J}$ and $\square$ : $B \simeq 4.2 \times 10^{-5}\;\rm{J}$. The images in the panels (below) are experimental snapshots corresponding to points (a)-(f) of the phase diagram (above).}
\label{fig4}
\end{figure}

 Varying $\dot{\gamma}$ and $B$ enable us to build a complete phase diagram (Fig.  \ref{fig4}). To build this diagram, we first reduced the number of parameters from four ($L$, $\dot{\gamma}$ and $B$  and $\Gamma$) to two non-dimensional ratios by using dimensional analysis.  
 As first non-dimensional parameter we choose the ratio of $L$ to  the characteristic length $L_c = (B/\eta \dot \gamma )^{1/3}.$  The length $L_c$ can be recognised as the buckling length of a sheet freely suspended in a shear flow away from boundaries \cite{salussolia2022simulation, kamal2021alignment,gravelle2025effect, verhille2022deformability, vaquero2026fluttering, miara2024folding} ($L/L_c$ is equivalent to the Sperm number used in  fiber hydrodynamics \cite{sulaiman2019numerical}). The second non-dimensional parameter is the ratio $L_{adh}/L_c $ comparing adhesion and viscous stresses, where  $L_{adh}$ is  the adhesion length $$L_{adh}= \Gamma / (\eta \dot \gamma).$$ 
 Notice that $L_{adh}/L_c$ is independent of $L$, and $L/L_c$ is independent of $\Gamma$.

Figure \ref{fig4} maps the transition between peeling and reattachment based on these two non-dimensional quantities, with red symbols indicating reattachment and green symbols indicating peeling. Despite some scatter in the data, the diagram seems to indicates a vertical boundary between the red and green region in the experimental range ($0.4<L_{adh}/L_c<20$). This indicates that for our experimental range the peeling/reattachment criterion is essentially set by  $L/L_c$ only. This observation  leads to the apparently paradoxical conclusion that adhesion, via the dependence on $L_{adh}$, is only marginally important in the current problem. How can this conclusion be rationalised? Consider the peeling threshold (\ref{eqn:criticalrate}). This prediction was obtained applying Griffith's theory to the stability of the interfacial fracture between an elastic sheet and a flat boundary, using a  simplified analytical model for the hydrodynamic load on the sheet, as explained in detail in Ref.  \cite{SALUSSOLIA2020103764} and briefly in Appendix B. Griffith's theory states  that the interfacial fracture will propagate when
\begin{eqnarray}\label{eqn:griffith}
\Gamma  =  - \frac{\partial U}{\partial L},
\end{eqnarray} 
where $U = E_b - W_H$ is the potential energy (per unit transversal length) of the non-adhered part of the sheet, $E_b$ is the bending energy and $W_H$ is the external work done by the hydrodynamic force on the flexible sheet, both per unit transversal length. The infinitesimal change $\frac{\partial U}{\partial L}$ of the total potential energy with respect to a change in debonded sheet length is the strain energy release rate (in non-dimensional form,  (\ref{eqn:griffith}) is $L_{adh}/L_c = -(L_c^2/B)(\partial U/\partial L)$). Ref. \cite{SALUSSOLIA2020103764} provides solutions for (\ref{eqn:griffith}) according to a simplified hydrodynamic model. The basic argument leading to eq.\eqref{eqn:criticalrate} from (\ref{eqn:griffith}) is that if ${\partial U}/{\partial L}$ diverges in a range of values near $L/L_c \sim 1$, then whatever the value of $\Gamma$ the instability condition \eqref{eqn:griffith} can be met for $L/L_c \sim 1$ independent of the specific value of  $\Gamma$. For example, if we consider two values of adhesion energy, say $\Gamma_1 = 0.4 N/m$ and $\Gamma_2 = 0.8 N/m$, the values of $L=L_2$ that match eq. \eqref{eqn:griffith}  for $\Gamma_2$ will only be very slightly larger than the value $L_1$ corresponding to $\Gamma_1$, if $U$ depends strongly on $L$ in the neighbour of $L_1 \simeq L_2$. 

\begin{figure}[h!]
\centerline{\includegraphics{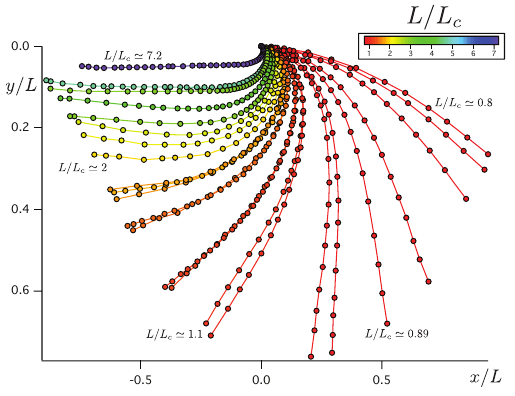}}
\caption{Equilibrium shapes for a clamped sheet for selected values of $L/L_c$. In  this experiment one end of the sheet is permanently adhered, so the inter-layer fracture cannot propagate .}
\label{fig3}
\end{figure}

We wanted to test the basic assumption leading to \eqref{eqn:criticalrate}, namely the approximate divergence of $U$ in the neighbourhood of $L=L_c$. The bending energy depends only on the configuration of the sheet, which we have experimental access to. Therefore, we examined how $E_b$  varies with $L$. We did this by considering a case where the peeling front is pinned. Experimentally, this is achieved by clamping the sheet at one end, leaving a length $L$ detached. We investigated the effect of $L_c = (\eta \dot \gamma /B)^{1/3}$ on the shape of the sheet  by considering combinations of $\dot \gamma$ and $L$ (Fig. \ref{fig3}). For $L \ll L_c$ , the sheet curvature is comparatively small. The flap is raised but is not vertical (the inclination angle of the flap is roughly $45^{\circ}$ for $L/L_c\simeq 0.8$). A vertical sheet is observed for  $L/L_c$ slightly larger than $\simeq 0.89$.  For $L/L_c \simeq 1.1$ the sheet points downstream and for $L/L_c\simeq 2$ the tip of the sheet is almost aligned with the undisturbed flow. For larger $L/L_c$ the sheet is curved only near the pinned front. The rest of the sheet is aligned with the undisturbed flow.  This is  the $\pi$-peel configuration investigated in Ref. \cite{botto2019toward}.

\begin{figure}[h!]
\centerline{\includegraphics{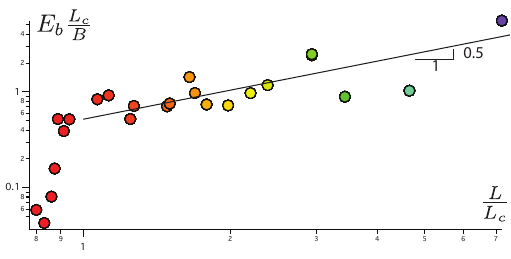}}
\caption{Bending energy versus $L/L_c$ for the shapes of fig.\ref{fig3}.}
\label{fig3ad}
\end{figure}

\begin{figure}[h!]
\centerline{\includegraphics{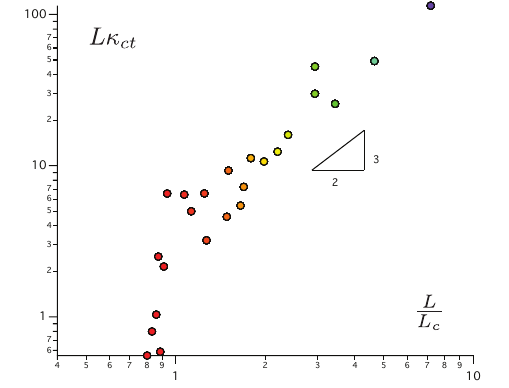}}
\caption{Sheet curvature at the peeling front  vs. $L/L_c$ for the shapes of fig.\ref{fig3}. }

\label{fig3ab}
\end{figure}

From the sheet geometry, we calculated the local curvature  $\dot\theta (s)$ and the bending energy $E_b = 1/2 B \int_0^L  \dot\theta^2 ds$, where $s$ is the curvilinear coordinate. The bending energy, plotted in fig.\ref{fig3ad}, shows a sharp increase for $L/L_c$ in the range [0.8, 1], i.e. in the range in which the sheet displacement becomes large. The increase in bending energy is sharp: $E_b$ increase by one order of magnitude over a variation of the rescaled length of only $20\%$.  For larger $L/L_c$, the bending energy varies more smoothly, approximately with a power law of exponent $0.5$. Another way to look at the data is to examine the curvature $\kappa_{ct}$  at the the pinned front vs $L/L_c$. Now the divergence of the curvature at a critical value of $L/L_c$ slightly smaller than 1  is even more evident (Fig.\ref{fig3ab}).  The asymptotic power-law exponent is now close to 1.5.

\begin{figure}[h!]
\centerline{\includegraphics{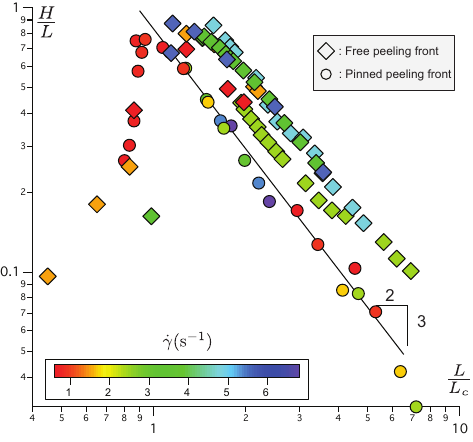}}
\caption{Normalised $y$ coordinate of the free tip of the sheet  versus $L/L_c$. Both data for clamped and non-clamped sheets are included. }
\label{fig3b}
\end{figure}
\begin{figure}[h!]
\centerline{\includegraphics{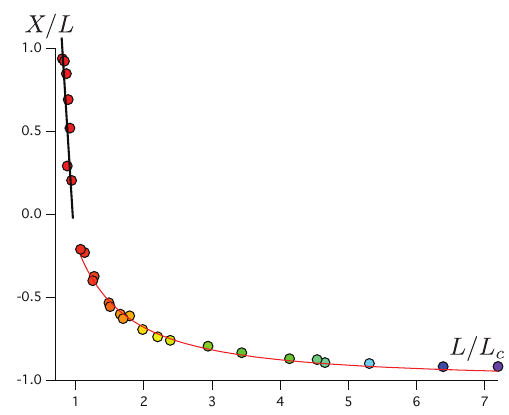}}
\caption{Normalised horizontal tip position $X/L$ versus $L/L_c$ for clamped sheets, corresponding to fig.\ref{fig3}. The black line equation is $X/L = a L/L_c + b $,  with $a \simeq -5.2$ and $b\simeq5.3$. The red line  is $X/L = -1 + A (L_c/L)^{\alpha}$, with $A\simeq0.82 $ and $\alpha\simeq 1.36$. }
\label{fig3ac}
\end{figure}

Building a peeling model requires a characterisation of the shape of the sheet (we have for example used this characterisation in the models contained in the Appendix). Furthermore, such characterisation can be used, e.g., for validation of numerical methods. Figure \ref{fig3b} shows the vertical position $H$ of the tip of the sheet, normalised by $L$.  An abrupt change in the range [0.8, 1] of $L/L_c$ is followed by a power law dependence $H/L  \sim  \left(L/L_c \right)^{-3/2}$ for larger $L/L_c$.  Figure \ref{fig3ac} shows the evolution of the horizontal rescaled position of the tip $X/L$ versus $L/L_c$. The variation in the range [0.8, 1] of $L/L_c$ is approximately linear. For $L/L_c>1$,  a sub-linear variation $X/L = -1 + A (L_c/L)^{\alpha}$ with $A\simeq0.82 $ and $\alpha\simeq 1.36$ fits the data reasonably well.

 The scaling laws observed for $L/L_c> 1$  can be explained by a  balance of bending moments.  For $L \gg L_c$ the total hydrodynamic force is essentially oriented along the flow direction  and is approximately  $ \eta \dot \gamma L$, the value expected for Couette flow past a plate of length $L$.  Because the height of the sheet is $H$, the moment of this hydrodynamic force is  $\sim \eta \dot \gamma L H$. This moment must balance the elastic moment  $\sim B/H$ resisting the deformation. Thus 
\begin{eqnarray}
\frac{H}{L}  \sim  \left(\frac{L_c}{L} \right)^{3/2},
\end{eqnarray}
The power law exponent $3/2$ is consistent with the experimental data of fig. \ref{fig3b}.  This scaling argument is also consistent with the sub-linear variation of $X/L$ observed in fig. \ref{fig3ac}. In the long sheet limit $L/L_c\gg1$, the length constrain reads $X \simeq - L + H$, which directly gives $X/L \simeq - 1 + (L_c/L)^{-3/2}$, close to the data in Fig. \ref{fig3ac}. The square root dependence of the bending energy seen in fig.\ref{fig3ad} follows from the same argument: for $L \gg L_c$, the tape is curved close to the peeling front, the typical curvature is $1/H$ and the length of the curved part is $\sim H$. Hence, the bending energy (per unit length) is $E_b \sim H B/H^2 \sim B/H \sim B \sqrt{L/L_c}/L_c$.

A power-law for the energy release rate $(L_c^2/B) (dU/dL) \sim (L/L_c)^{n}$  yields a peeling criterion $L_{adh}/L_c\sim (L/L_c)^{n}$ hence  $\eta \dot \gamma \sim B^{(n-1)/(n+2)} L^{-3n/(n+2)} \Gamma^{3/(n+2)} $. A fit of the experimental data to $E_b \sim (L/L_c)^{(n+1)}$ gives $n \simeq 9$ for $L/L_c$ around $1$. Assuming that $E_b$ follows approximately the same behaviour of $U$, we can see that the dependence on $\Gamma$ is indeed very weak, going a approximately as $\Gamma^{1/4}$  in our parameter range. For example, if the adhesion energy changes by a factor of $20$, which would correspond to a drastic change in the adhesion properties, the critical shear rate would only change by a factor of $2.1$.   This again supports our argument that the effect of adhesion, while not completely negligible, is markedly less important than the effect of bending in setting the value of the critical peeling shear rate.  

In figure \ref{fig3b} the clamped and freely moving sheet cases are compared for the variation of $H/L$. Despite some differences - the freely moving case is slightly larger than the pinned case - the dependence  of $H/L$ with the  normalised length $L/L_c$ is similar in the two cases. In particular, both cases follow approximately the same power law with an exponent $3/2$ for $L/L_c>1$. This  comparison indicates that in the dynamic case the shapes can be considered, to a first approximation, essentially quasi-static. 

Is hydrodynamics unimportant? Certainly not very near the peeling front. In this region, viscous forces are large and determine the rate at which the peeling front moves. While we were unable to develop a model for the case in which the sheet peels off under the effect of the applied shear,  we could model the reattachment dynamics using a model first proposed  to describe experiments on the reattachment in air of a thin silicon wafer. \cite{rieutord2005dynamics}A primary difference of our experiments with theirs is that in their case the shape of the sheet near the peeling front was invariant to translation. In contrast in our case the curvature of the sheet at the contact line decreases during reattachment, so the validity of the model is not obvious. 

The model uses the lubrication equation for the flow in the gap between $y=0$ and $y=h(x,t)$, and is based on equating the mechanical power dissipated in viscous friction to the energy gain $\Gamma V$ upon reattachment, where $V$ is the peeling front velocity. To avoid the well-known stress divergence at the contact line, the lubrication equation is assumed valid until $h$ becomes comparable to an assigned nanometric cut-off length scale $\ell_{cut}$. The scaling for $V$ predicted by this model is 
\begin{equation}
\label{eqn:velocity}
V \sim  \frac{ \Gamma^{5/4} \ell_{cut}^{1/2}}{B^{1/4} \eta }. 
\end{equation}
The  expression  gives results comparable with the experiments using $\ell_{cut} \simeq 1nm$ (model prediction $V \simeq 4.5 \times 10^{-6} {m.s^{-1}}$, experimental value $\simeq 3.4\times 10^{-6}\;\rm{m.s^{-1}}$,  see Appendix). In the more general case in which the sheet motion is driven by the shear flow, one would have to account for the power of the external force in addition to $\Gamma V$, but this is difficult owing to the complex geometry of the moving sheet. The peeling velocity scale (\ref{eqn:velocity}) can be compared to the velocity scale $\dot{\gamma} L$ for unhinged rotation of a plate of length $L$ \cite{pozrikidis2006flipping}. We find that the ratio of $V$ to $\dot{\gamma} L$ is much less than 1, compatibly with our observation of quasi-static deformation of the sheet.

When $L/L_c$  is large, the sheet is subject to an applied force per unit width $ \sim \eta \dot{\gamma}L$, directed almost parallel to the flow. The linearity of Stokes flow requires the peeling front velocity $V$  to be linear in the force, provided that  the shape of the sheet in the vicinity of the contact line is independent of the applied force. Thus, perhaps not unexpectedly, we observe asymptotically a linear scaling between $V$ and $L$ (Fig.\ref{figA}).

\begin{figure}
\centerline{\includegraphics[width = 9cm]{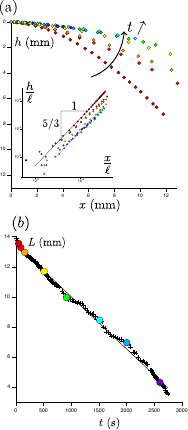}}
\caption{ Re-attachment dynamics in quiescent flow for a tape of bending modulus $B\simeq 6.9\times10^{-7}  J$. (a) Shapes of the detached part of the tape during the re-attachment. The caption presents the same data rescaled with the length $\ell  = \sqrt{B/\eta  V}\simeq 0.33\;\rm{m}$. The colors of the markers of the different shapes of panel (a) correspond to the selected times of panel (b). (b) Evolution of the detached length versus time. From this graph, we calculate a constant front propagation velocity $V\simeq 3.4\times 10^{-6}\;\rm{m.s^{-1}}$. The colors in (b) correspond to those in (a).}
\label{fig2}
\end{figure}

\section{Discussion: which peeling formula?}
\label{sec:discussion}

In our experiments we have observed a critical peeling shear rate $\gamma  \sim B/\eta L^3$ . This formula is valid for an experimental range, and is part of a set of peeling formulas which are valid in different regimes of adhesion.  Based on the analysis reported in the Appendix and which recapitulate some results of Ref. \cite{SALUSSOLIA2020103764}, three different formulas, corresponding to three different peeling regimes, have been identified. To illustrate the use of these peeling formulas, we can consider two practical cases where the results of this paper will be applied: i) experiments where the value of the imposed shear rate is given and the calculation of the minimum (critical) length $L$ for peel off is required; and ii) experiments where the debonded length is given, and the calculation of the critical $\dot{\gamma}$  is required. 

Let's examine case i) first. The parameters $\eta \dot \gamma$, $\Gamma$ and $B$ are fixed. The length scales $L_c= (B/\eta \dot \gamma)^{1/3}$ and $L_{adh} = \Gamma / \eta \dot \gamma$ are therefore fixed. If $L_{adh} \ll L_c$, the peeling front is downstream of the tip of the flap (the flap aperture is small). The critical peeling length is $L=(L_{adh} L_c^3)^{1/4}$, which correspond to $\eta \dot \gamma = \sqrt{\Gamma B}/L^2$. This limit corresponds to the regime of peeling at constant load, as can be recognised by the square root dependence of $\dot{\gamma}$ on $\Gamma B$\cite{roman2013fracture}. If $L_{adh} \gg L_c$, the peeling  length is $L=L_{adh}$, corresponding to $\eta \dot \gamma = \Gamma/L$. In this case the tip of the flap is well downstream of the peeling front, $L \gg L_c$. If $L_{adh} \sim L_c$, the peeling length is $L=L_c$, hence $\eta \dot \gamma \sim B/L^3$.  To summarise:

\begin{eqnarray}
\text{Weak adhesion/high rigidity: } L_{adh} \ll L_c : L \sim \frac{(\Gamma B)^{1/4}}{ \sqrt{\eta \dot \gamma}} \\
\text{Intermediate adhesion/rigidity: } L_{adh} \sim L_c  \sim 1  : L \sim \left(\frac{ B}{\eta \dot \gamma} \right)^{1/3}  .\\
\text{Strong adhesion/low rigidity: } L_{adh} \gg L_c   : L  \sim \frac{\Gamma}{\eta \dot \gamma } .
\end{eqnarray}

In case ii), $L$, $\Gamma$ and $B$ are fixed. The  three adhesion regimes are :
\begin{eqnarray}
\text{Weak adhesion/high rigidity:} \frac{\Gamma L^2}{B} \ll 1  : \eta \dot \gamma \sim \frac{\sqrt{\Gamma B}}{L^2} \\
\text{Intermediate adhesion/rigidity:} \frac{\Gamma L^2}{B} \sim 1  : \eta \dot \gamma  \sim B/L^3.\\
\text{Strong adhesion/low rigidity: } \frac{\Gamma L^2}{B} \gg 1  : \eta \dot \gamma  \sim \Gamma/L .
\end{eqnarray}
Our experiments are in the intermediate adhesion/rigidity regime. Notice that as the peeling front moves $L$ increases. Thus from the intermediate adhesion regime a peeled sheet  will eventually enter the strong adhesion regime. But because $B/L^3$ is larger than $\Gamma/L$ when $\frac{\Gamma L^2}{B} \gg 1$ the motion of the peeling front does not stop.

\section{Conclusions}
We studied experimentally the viscous peeling of a flexible sheet adhered to a flat substrate. The peeling process is triggered by a steady linear shear flow. The results demonstrate that the observed peeling threshold is of the order of $\dot \gamma \sim \frac{B}{\eta L^3}$, where $L$ is the length of the debonded portion of the sheet,  $B$ is the bending rigidity, and $\eta$ is the fluid viscosity.  For values smaller than this threshold, the sheet re-adheres to the substrate, with a velocity that we are able to predict. Our results are valid in the limit of intermediate bending and intermediate adhesion, as defined in Sec. \ref{sec:discussion}.  We explain the weak dependence of the critical shear rate for peeling on $\dot \gamma$ by demonstrating a strong dependence of the sheet's curvature on $L/L_c$ when $L$ is close to $L_c$. Our experimental findings confirm a theoretical hypothesis put forward in  Ref. \cite{SALUSSOLIA2020103764}.

According to the model, for $L \rightarrow 0$ the value of $\dot{\gamma}$ required to remove the sheet goes to infinity. While this is a mathematical idealisation, we indeed find that sheets for small values of $L$ the sheet is extremely difficult to remove even increasing $\dot{\gamma}$. It would be interesting to repeat our experiment with unsteady flows, to examine the possibility that unsteady velocity fluctuations promote detachment for small $L$. To simplify the already challenging experiments, sheets clamped at one edge via, for example, a thin strip of permanent adhesive  could be used.    

From a fluid dynamical perspective, the precise quantification of the rate of work done by the hydrodynamic force would be useful to characterize the rate of peeling.  Numerical methods for fluid-structure interactions are now well able to simulate large scale non-linear deformations. These should be matched with analytical models to avoid the stress singularity at the peeling front. \cite{sui2014numerical} There is also space for analytical work: the flow in the middle of the wedge, on a scale  $ \ell \ll w$, is essentially two-dimensional, while the flow outside the flap, on a scale $ \ell \gg w$, is fully three-dimensional.  An application that calls for the methods of matched asymptotic expansions.

From a practical perspective, our work could provide an explanation for a discrepancy observed in the analysis of  shear-induced exfoliation of graphite into graphene. Molecular dynamics simulations of  delamination of nanoscopic graphite particles show that the dominant exfoliation model is sliding, where the layers slide off past each other without bending.\cite{JCPgravelle20}  A model balancing the work of the viscous stress and the adhesion energy \cite{JCPgravelle20} and giving a critical shear rate $ \dot \gamma \sim \Gamma/ L \eta$ matched the molecular dynamics data perfectly, provided that $\Gamma$ was estimated correctly and a slip correction accounting for the incomplete adherence of the fluid to the solid was considered.   On the other hand the sliding mechanism is not able to predict the experimentally observed critical shear rate of $\dot \gamma \simeq10^{4}\rm{s^{-1}}$ to exfoliate typical graphite particles of $L\simeq 1\;\rm{\mu m}$.  \cite{Paton2014ty}  One explanation for this disagreement could be the importance of graphene bending, which would promote detachment through a peeling mechanism similar to the one we observe in our experiments. \cite{salussolia2022simulation, qi2025unraveling}

We believe that the current work could have implications for many high-tech areas that are now emerging and that require an understanding of the coupling between thin sheets, adhesion, and moving liquids and gases. Besides the afore-mentioned fluid-assisted delamination of 2D materials \cite{botto2019toward, li2020mechanisms, yi2015review} we can mention  wet transfer of 2D materials from growth substrate to devices \cite{chen2016progress, pyo2025etchant}. Our results could also give indications for particulate removal in a variety of clean-room applications where particulate contamination must be kept to negligible levels.  \cite{van2019advanced, kilroy2011adhesion}A critical insight from the current paper is that estimates based on the balance ``adhesion=work by drag force'' may lead to vastly incorrect predictions when the foreign particle has a height much smaller than the lateral width, because this balance fails to incorporate the effect of the body's deformation on the force resisting detachment. We have demonstrated that for a thin particulate sheet having $L$  close to $L_c$, the change in elastic bending deformation of the body is large upon even a small increment in debonded length, with a large effect on the force of detachment.


\section*{Author contributions}
L.B. supervised the research; P.H. and L.B. designed research; P.H. performed experiments; P.H. and L.B. analyzed data; P.H. and L.B. wrote the paper.

\section*{Conflicts of interest}
There are no conflicts to declare.

\section*{Data availability}
Data are available at TU Delft's data repository upon request to the authors.

\section*{Acknowledgements}
We gratefully acknowledge funding by the European Research Council (ERC) under the European Unions Horizon 2020 Research and Innovation program (Project FLEXNANOFLOW, Grant No. 715475).

\section{Appendix A :Dynamics in quiescent flow}
 An adhered sheet of  bending modulus $B\simeq 6.9\times10^{-7}  J$  with an initial detached length $L\simeq 1.4\;\rm{cm}$ was let to relax without imposing a shear flow. The tape reattaches on the flat substrate as seen on the profiles for selected times on fig.\ref{fig2}(a). Fig.\ref{fig2} (b) shows the detached length versus time. 

The reattachment is seen to occur at a constant velocity $V\simeq 3.4\times 10^{-6}\;\rm{m.s^{-1}}$. To model $V$ we consider the balance between viscous, elastic and adhesion power in the lubrication approximation. The lubrication flow velocity and the pressure gradient in the gap between the reattaching flap and the flat substrate are given by 

\begin{eqnarray}
 u_x(x,y) &=&   6  V\frac{  (h-y) ( \ell_{cut}+y)}{h (h + 3  \ell_{cut})} \\
 \frac{\partial P}{\partial x} &=& - V   \frac{12 \eta} {h (h + 3  \ell_{cut})}
\end{eqnarray}
where we have introduced a cut off length $\ell_{cut}$ to avoid the well-known stress singularity at the peeling front \cite{sui2014numerical}.
Neglecting tangential hydrodynamic stress, the linear Euler-Bernoulli equation governing the shape of the sheet is 
\begin{eqnarray}
B\frac{\partial^4 h}{\partial x^4} &=&P
\end{eqnarray}
We use the length $\ell  = \sqrt{\frac{B}{\eta  V}}\simeq 0.33\;\rm{m}$ to rescale the Euler-Bernoulli equation obtaining
\begin{eqnarray}
\frac{\partial^5 \tilde h}{\partial \tilde x^5} =-  \frac{12} {\tilde h (\tilde h + 3  \tilde \ell_{cut})}
\end{eqnarray}
where the ``tilde'' sign denotes rescaled variables. Away from the reattachment front, we have $\tilde h \gg \tilde \ell_{cut}$ thus
$\frac{\partial^5 \tilde h}{\partial \tilde x^5} =  -\frac{12} {\tilde h^2}$.
This equation admits a power-law solution $\tilde h(\tilde x)= A \tilde x^\alpha$ if  the exponent verifies $\alpha -5 = -2 \alpha$ leading to $\alpha = 5/3$ and if the prefactor $A$ verfies  $(\alpha -4) (\alpha -3) (\alpha -2) (\alpha -1) \alpha  A = -12 A^{-2}$ leading to $A\simeq 2.18$.
The experimental data of figure \ref{fig2} shows the sheet shapes during the reattachment dynamics. It confirms that the shapes are well described by a power law of exponent $5/3$ (although we find from the experimental values that the prefactor $A$ varies weakly on time, from $5.6$ to $1.7$).

The rate of viscous dissipation in the wedge is 
  \begin{eqnarray}
 \Phi  &=&  \int_0^h \int_0^{L}\frac{\partial P}{\partial x} u_x dx dy \\
& \simeq&   \frac{12  }{5 A^{3/5}} \left(\frac{\ell}{ \ell_{cut}}\right)^{2/5}\eta  U^2 = \frac{12  }{5 A^{3/5}}  \frac{\sqrt[5]{B} \eta ^{4/5} U^{9/5}}{ \ell_{cut}^{2/5}}
\end{eqnarray}
Balancing $\Phi$ to the rate of work done by adhesive forces $\Gamma V$ gives 
\begin{equation} 
V  = \frac{5 }{12}A^{3/5} \frac{ \Gamma }{  \eta } \left(\frac{ \ell_{cut}}{ \ell} \right)^{2/5} 
\end{equation} 
which can also be rewritten as
\begin{equation}
  V = \frac{5 }{12\sqrt{2} } \sqrt[4]{\frac{5}{3}}A^{3/5}  \frac{ \Gamma^{5/4} \ell_{cut}^{1/2}}{B^{1/4} \eta }   \end{equation} 
  For realistic values $B\simeq 6.9\times10^{-7}  J$, $\Gamma \simeq 0.03 N/m$, $A\simeq 1$, $\eta \simeq 1 Pa.s$, $\ell_{cut}\simeq10^{-9}m$ we get $V \simeq 4.5 \times 10^{-6} {m.s^{-1}}$, close in order of magnitude to the the experimental value $\simeq 3.4\times 10^{-6}\;\rm{m.s^{-1}}$.

%

\section{Appendix B : Analysis of peeling regimes}

\textbf{\textit{Weak adhesion, high rigidity regime:  $L \ll L_c$ and $L_{adh} \ll L_c$}}

The regime $L \ll L_c$ and $L_{adh} \ll L_c$ is a weak adhesion, high rigidity  regime, because the rescaled adhesion $ \Gamma L^2/B = L_{adh} L^2 / L_c^3  \ll 1$.

The sheet is only slightly bent. For small tip opening angles, the fluid below the detached sheet is practically quiescent, so the pressure load acting on the debonded portion of the sheet is approximately uniform and proportional to $ \eta \dot \gamma $\cite{SALUSSOLIA2020103764}. This gives $U \propto {\dot \gamma} ^2 $, and the critical shear rate depends on the square root of $\Gamma$ \cite{SALUSSOLIA2020103764, botto2019toward, barbieri2022peeling}:  
\begin{eqnarray}
\dot \gamma \sim \frac{\sqrt{\Gamma B}}{\eta  L^2}
\label{shearrate2}
\end{eqnarray}
In terms of $L/L_c$ this expression corresponds to  $-\frac{L^2_c}{B}\frac{dU}{dL} \propto (\frac{L}{L_c})^4$. 

\textbf{\textit{Strong adhesion, low rigidity regime:  $L \gg L_c$ and $L_{adh} \gg L_c$}}

The regime $L \gg L_c$  and $L_{adh} \gg L_c$  is a strong adhesion , low rigidity  regime, because the rescaled adhesion $ \Gamma L^2/B  = L_{adh} L^2 / L_c^3  \gg 1$. 

In this regime the flap is practically parallel to the substrate and is subject to an axial tension $\eta \dot{\gamma}L $. In this case the work done by the shear stress balances adhesion exactly \cite{roman2013fracture}:

\begin{equation}
\dot \gamma \sim \frac{\Gamma}{\eta L}
\end{equation}

\begin{figure}
\centerline{\includegraphics{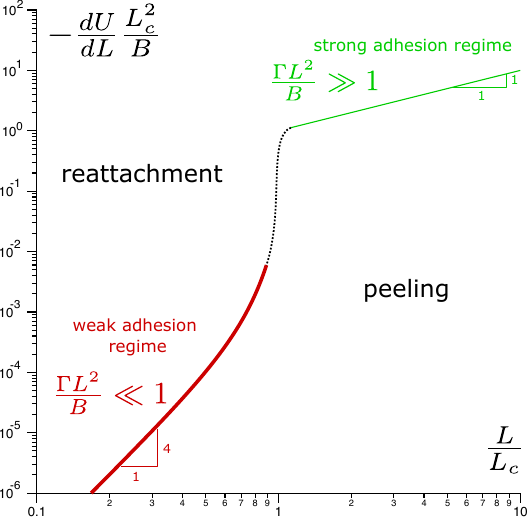}}
\caption{ Dependence of the normalised strain energy release on $L/L_c$. The continuous lines correspond to equation \eqref{pc2} (red) and equation \eqref{pc1} (green). The limit $L/L_c \rightarrow 0$ of Eq. \eqref{pc2} gives $-dU/dL(L_c^2/B) \propto (L/L_c)^4$, as in the weak adhesion regime (see paragraph before Eq. $\ref{shearrate2}$). For $L/L_c$ slightly smaller than 1, Eq. \eqref{pc2} models the first effect of the angle-dependent hydrodynamic load on the strain energy release rate.  The dashed line indicates a possible smooth interpolation (in the medium adhesion-rigidity regime) between the red and green lines.  }
\label{fignum}
\end{figure}

A more accurate model can be derived based on the results of the current paper.  We have seen that for $L \gg L_c$ the horizontal position of the tip of the sheet follows approximately $X/L = -1 + A (L_c/L)^{\alpha}$ (see fig.\ref{fig3ac}). For a variation $dL$, the tip moves by $ (- dL + dX)$ in the $x$ direction. The infinitesimal work done by the hydrodynamic force is $ dW_H =  \eta \dot \gamma L (dL - dX)$ and thus 
\begin{eqnarray}
\frac{ \partial W_H}{ \partial L} =   \eta \dot \gamma L \left(  1 - \frac{d X}{dL} \right) \sim  \eta \dot \gamma L \left( 2 + A (\alpha -1) \left(\frac{L_c}{L}\right)^{\alpha} \right).
\end{eqnarray}
The typical sheet curvature near the peeling front is $1/H$ and the length of the curved part is $\sim H$. Hence the bending energy $E_b \sim H B/H^2 \sim B/H$. The variation of the bending energy is
\begin{eqnarray}
\frac{dE_b}{dL} = B \frac{d (1/H)}{dL} = B \frac{1}{L_c \sqrt{L L_c}}
\end{eqnarray}
and thus 
\begin{eqnarray}\label{pc1}
-\frac{\partial U}{\partial L} =   \eta \dot \gamma L \left( 2 + A (\alpha -1) \left(\frac{L_c}{L}\right)^{\alpha} \right)  -  B \frac{1}{L_c \sqrt{L L_c}}
\end{eqnarray}
This last expression is plotted, in dimensionless form, in fig.\ref{fignum} for $L/L_c >2 $ (see the green curve). For $L/L_c \gg 1$, $-\frac{dU}{dL}$ follows a power law scaling of exponent $\simeq 1$. Using expression \textcolor{red}, this scaling translates to  
\begin{eqnarray}
\dot \gamma = \frac{ \Gamma}{2 \eta  L}. 
\label{shearrate1}
\end{eqnarray}

\textbf{\textit{Intermediate adhesion, intermedia rigidity regime:  $L \sim L_c$ and $L_{adh} \sim L_c$}} An analytically tractable model for this regime can be obtain by noting that the pressure acting on the flap should depend on the opening angle of the flap. Indeed, as the opening angle increases a larger portion of the flap is exposed to a free-stream flow having increasing velocity.  In Ref. \cite{SALUSSOLIA2020103764}, this effect was modelled by assuming a pressure variation of the form $\mu \dot \gamma (q_0 + q_1 \theta)$, where $\theta$ is the opening angle at the tip.  The analytical solution for this angle- dependent hydrodynamic load is 
\begin{eqnarray}\label{pc2}
- \frac{dU}{dL} = \frac{B}{L_c^2}\left(\frac{L}{L_c}\right)^4  \frac{8q_0^2\left(\left(\frac{L}{L_c}\right)^3
   q_1+40\right)}{5 \left(8- \left( \frac{L}{L_c}\right)^3 q_1\right)^3},\end{eqnarray} 
where $q_0 \simeq 0.1$ and $q_1 \simeq 5.37$. These model parameters were extracted from high-resolution simulation of the incompressible Stokes equation calculations of a sheet having a simplified wedge-like geometry with a uniform opening angle $\theta$ from the tip to the peeling front \cite{SALUSSOLIA2020103764}.  Expression  \eqref{pc2} diverges as $L/L_c$ approaches $(8/q_1)^{1/3}$ and recovers (\ref{shearrate2}) when $L\ll L_c$. While Eq. \eqref{pc2} does not  capture the full non-linear deformation of the flexible sheet, and the complex dependence of the local hydrodynamic force on the sheet's shape, it does account in an effective manner on the opening of the sheet via the angle $\theta$.


The dependence of the normalised strain energy release rate on $L/L_c$ in the three regimes identified is illustrated in fig.\ref{fignum}. A connecting black dashed line is plotted in correspondence to the regime of intermediate adhesion and rigidity to illustrate a possible smooth interpolation between the asymptotic expressions \eqref{pc2} (continuous red line) and equation \eqref{pc1} (continuus green line). It is seen from this plot that for $L/L_c$ varying in the narrow range $[0.8,2]$,  $(dU/dL) (L_c^2/B)$  varies by approximately 3 orders of magnitude, a very strong dependence.

\balance


\bibliography{biblio.bib} 
\bibliographystyle{rsc} 

\end{document}